ARTICLE TYPE

# World of ScoreCraft: Novel Multi-Scorer Experiment on the Impact of a Decision Support System in Sleep Staging

Benedikt Holm[1] | Arnar Óskarsson[1] | Björn Elvar Þorleifsson[1] | Hörður Þór Hafsteinsson[1] | Sigríður Sigurðardóttir[2] | Heiður Grétarsdóttir[2] | Kenan Hoelke[2] | Gabriel Marc Marie Jouan[1] | Thomas Penzel[4] | Erna Sif Arnardottir[2,3] | María Óskarsdóttir[1]

[1]School of Technology, Department of Computer Science, Reykjavik University, Iceland

[2]School of Technology, Sleep Institute, Reykjavik University, Iceland

[3]The National University Hospital of Iceland, Landspitali, Iceland

[4]Sleep Medicine Center, Charite - Universitätsmedizin Berlin, Berlin, Germany

**Correspondence**
Corresponding author Benedikt Holm,
Email: benedikthth@ru.is

Funding Information
This research was supported by the European Union's Horizon 2020 research and innovation programme (grant agreement 965417). Holm B. received funding from the ErasmusPlus programme to travel to Penzel's lab in Berlin.

**Abstract**

Manual scoring of polysomnography (PSG) is a time-intensive task, prone to inter-scorer variability that can impact diagnostic reliability. This study investigates the integration of decision support systems (DSS) into PSG scoring workflows, focusing on their effects on accuracy, scoring time, and potential biases toward recommendations from artificial intelligence (AI) compared to human-generated recommendations.

Using a novel online scoring platform, we conducted a repeated-measures study with sleep technologists, who scored traditional and self-applied PSGs. Participants were occasionally presented with recommendations labeled as either human- or AI-generated.

We found that traditional PSGs tended to be scored slightly more accurately than self-applied PSGs, but this difference was not statistically significant. Correct recommendations significantly improved scoring accuracy for both PSG types, while incorrect recommendations reduced accuracy. No significant bias was observed toward or against AI-generated recommendations compared to human-generated recommendations.

These findings highlight the potential of AI to enhance PSG scoring reliability. However, ensuring the accuracy of AI outputs is critical to maximizing its benefits. Future research should explore the long-term impacts of DSS on scoring workflows and strategies for integrating AI in clinical practice.

**KEYWORDS**
Decision support system, Artificial intelligence, Sleep staging, Scoring accuracy

## 1 | INTRODUCTION

As sleep disorders and sleep issues are extremely prevalent in society (Arnardottir et al. 2016, Adams et al. 2017, Zeng et al. 2020), sleep technologists are more important today than ever. The sleep technologist is responsible for reviewing sleep studies, or polysomnograms (PSG), and manually annotating (also known as scoring) sleep stages and events such as movement, arousal, and apnea (American Academy of Sleep Medicine 2023). Medical doctors subsequently use this analysis for diagnosis.

A PSG is an overnight collection of biometric signals, such as various respiratory signals, electrooculogram (EOG), electroencephalogram (EEG), and electromyogram (EMG), to name a few (American Academy of Sleep Medicine 2023). The scoring of a PSG is a time-consuming task, which takes up to two hours to score a single eight-hour PSG (Rayan et al. 2023), and can significantly strain the sleep technologist and induce scoring fatigue. Alongside being laborious and time-consuming, the scoring of a PSG can also vary significantly between sleep technologists, with disagreement on sleep staging being as high as 14% (Nikkonen et al. 2024) and respiratory events as high as 34.6% (Rosenberg and Van Hout 2014), and obstructive apnea severity classifications differing in 66% of cases depending on the scorer (Pitkänen et al. 2024).

**Abbreviations:** AI, artificial intelligence; DSS, decision support system;

To reduce the need for a sleep technologist to set up and monitor a traditional PSG and to eliminate the requirement to sleep in a sleep laboratory overnight, new PSG equipment has been developed for home use. This multichannel frontal PSG enables patients to apply the device themselves and sleep comfortably in their own beds. These self-applied PSG devices are currently undergoing validation and a recent device demonstrated a success rate of approximately 90% in enabling effective self-application and accurate data collection (Ferretti et al. 2022).

The rise of artificial intelligence (AI) has not left the scoring process behind, with various automatic algorithms being designed to automate and speed up the work of sleep technologists by detecting sleep stages (Wara et al. 2024) and apnea (Moridian et al. 2022, Mostafa et al. 2019) to name a few tasks. These algorithms have great potential to assist in the scoring process; however, they cannot be treated as a drop-in replacement for the human expert since they are incapable of adapting and adjusting to the evolving standards of care as the human experts are. They also need to be rigorously trained for different issues that may arise in diverse application contexts and be aware of possible biases that occur in AI based on sex, age, comorbidities, and other factors (P?S? 2023).

To meet the need for systems that aim to utilize AI as a tool for the human expert instead of as a replacement, decision support systems (DSS) have been designed to provide interactive tool sets to assist experts in making decisions and solving unstructured or semi-structured tasks (Sprague 1980). Garg et al. (2005) found that in 64% of cases, DSS or similar systems considerably impacted clinician performance. Articles on DSS have been widely written in the field of sleep research; however, the literature covers mostly automation of individual tasks, but does not examine the effect of the inclusion of DSS into the workflows of sleep technologists in terms of accuracy or time taken to score sleep recordings. Furthermore, the effects of integrating DSS and AI into sleep technologists' workflows can provide significant advantages in terms of speed and accuracy, but to be integrated effectively requires building trust towards the AI (Asan et al. 2020).

Along with the aforementioned effects, an important aspect to measure is the potential impact of the DSS in changing the behavior of the human expert or how the professionals whose toolset is augmented with AI might, for example, become complacent and default to the AI recommendations (Parasuraman and Manzey 2010). Complacency towards AI can be defined as a tendency of the user to not appropriately scrutinize the results of the automated tools. The tendency towards complacency is complicated to analyze but has been shown to be linked to the transparency of the AI system, as well as how well the expert expects the AI to perform (Harbarth et al. 2024). Another important aspect of integrating AI is the concept of 'clinical acquiescence,' defined by Holm et al. (2024), which refers to the willingness to adopt AI assistance in clinical workflows.

There is a noticeable gap in the literature on the effects of integrating DSS into the workflows of sleep technologists. Most of the existing literature focuses on the accuracy of the algorithms designed to automate sleep-scoring tasks (Rusanen et al. 2024, Holm et al. 2024). However, the impact assessment of such algorithms on expert performances is a key component that is usually missing. We propose to leverage this by studying the changes introduced by human experts whose toolsets have been augmented with DSS. In more detail, we aim to investigate the effects of integrating AI into the work environment.

This work investigates the effects of introducing recommendations in scoring sleep stages. We further measure the effects of the recommendation presentation and correctness on the accuracy and speed of the human expert.

This study used a repeated measures design with two conditions to collect a consensus scoring for one hour of traditional and self-applied PSG. The main conditions being researched were the effect of recommendations presence and study type, as the objective of this study is to research the effects of recommendations on the sleep staging process. The study also examines the effect of the type of scoring recommendations (human or AI) on the scoring process, counterbalancing the recommendations by only showing recommendations for one session of each PSG type. Hence, 3 factors were being studied: type of sleep study, presence of recommendations, and type of recommendations. Their significance is measured in terms of scoring accuracy or correctness and time.

## 2 | METHODOLOGY

For this study, the scoring sessions were limited to a single hour (120 epochs), chosen from a data set that was recorded simultaneously using traditional PSG and self-applied PSG equipment. The hour was chosen for good-quality signals, and the hypnogram featured multiple transitions between sleep stages. The sessions were limited to one hour to focus on collecting a greater diversity of scorings and to prevent scoring fatigue from affecting sleep technologists during the process.

### 2.1 | Platform

The scoring collection was performed with the MicroNyx online scoring platform (Holm 2023), which allows secure online scoring of PSGs. The MicroNyx platform enables measuring difficult-to-obtain features, such as the decision-making time, change of mind, and more aspects of the scoring process. In preparation for this study, multiple sleep technologists were recruited before the experiment to validate and provide recommendations and feedback on the scoring interface and signal filtering to measure the impact of the new MicroNyx system on the scoring in a Co-Design process. This was repeated several times over a few months until each sleep technologist at Reykjavik University could reliably score with 80% consensus with a pre-existing scoring, or roughly around the 86% expected agreement of sleep technologists (Nikkonen et al. 2024). MicroNyx allows for the creation and the scoring of so-called 'scoring sessions,' variable-length signal segments containing signals that sleep technologists can then score for research purposes.

The signals, filtering, and data source can be customized and tailored to different research purposes.

## 2.2 | Recommendations

In 50% of epochs, participants were presented with a recommendation for a sleep stage. The recommendation rate of 50% was chosen to provide an equal amount of epochs with and without scoring recommendations. To investigate the presence of any potential bias against automatic algorithms, the recommendations were sourced from a human scoring for the PSG and self-applied PSG studies, respectively. For consistency, the same sleep technologist performed the scoring for the PSG and self-applied PSG recommendations (blinded to which recordings were from the same participants). Despite all scoring being sourced from the same human sleep technologist, the recommendations were either presented as being from a human sleep technologist or an automatic AI scoring system. This is henceforth referred to as the recommendation presentation. The ratio of human vs. AI recommendations was equal to ensure equal representation.

Both accurate and deliberately incorrect recommendations were implemented to evaluate the influence of recommendations on scoring behavior. The rate of incorrect recommendations was set to 20%, aligning with the inter-scorer variability reported by Nikkonen et al. (2024) and emulating the error rate of sleep technologists. To ensure realistic incorrect recommendations, a sleep-stage map was developed using results from a multicentric consensus scoring study (Nikkonen et al. 2024), which identified the most common misclassifications by sleep technologists. Figure 1 illustrates the most common misclassifications, providing insight into how incorrect recommendations were designed to reflect typical scoring disagreements.

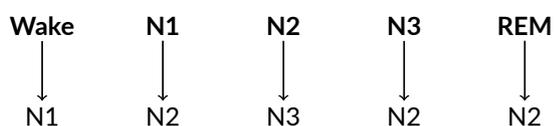

**FIGURE 1** Map of sleep stages to deliberate misclassification for scoring recommendations.

Recommendations presented as being from humans were represented with a scientist emoji (see Figure 2a). In contrast, the recommendations we presented as being from an AI had the robot emoji, as shown in Figure 2b. The recommendations were designed to be noticeable for sleep technologists without being intrusive or obscuring the signals meaningfully. The post-study survey included questions about the visibility of the recommendations to ensure that the participating sleep technologists could easily spot the recommendations and tell the difference between human and AI recommendations.

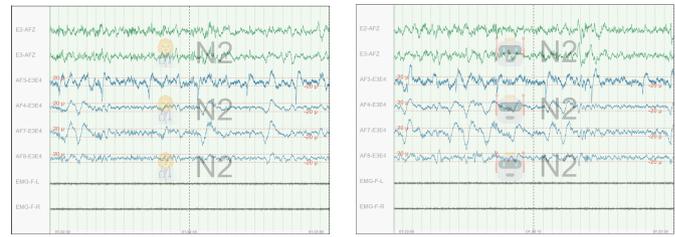

(a) Recommendation presented to be from a human    (b) Recommendation presented to be from an AI

**FIGURE 2** Comparison of recommendations presented as human vs artificial intelligence (AI)

## 2.3 | Data setup

The data for this study was selected from a 'double-setup' dataset, where a traditional PSG recording along with a self-applied PSG setup was placed on the participants, and the two types of PSG were recorded simultaneously (Rusanen et al. 2024).

The sleep technologists scored both types of PSG using the MicroNyx web scoring interface, which supports flexible and customizable signal selection and filtering. Scoring guidelines for required signals and their appropriate filtering were followed to ensure consistency with established methodologies and software for both traditional and self-applied PSG. This subsection is divided into two parts, detailing the signals and filtering options for traditional PSG and self-applied PSG, respectively.

### Traditional PSG

The traditional PSG setup presented to the sleep technologists was based directly on the AASM recommendations for scoring sleep stages, utilizing the appropriate signals and filters as listed in the guidelines (American Academy of Sleep Medicine 2023). An illustration of the PSG signals is provided below.

The signals the sleep technologists used to score the traditional PSG were the EEG signals C4-M1, C3-M2, F4-M1, F3-M2, O1-M2, O2-M1, the EOG signals, E1-M2, E2-M1, and the chin EMG.

The EEG signals were filtered using a 0.5–35 Hz bandpass filter. Each EOG signal was processed through a 0.3 -35 Hz bandpass filter and sampled at 200 Hz. The chin EMG signal was passed through a 10 Hz high-pass filter and sampled at 200 Hz.

Figure 3 shows how the signals from the traditional PSG were presented to the sleep technologists in the MicroNyx scoring interface.

### Self-applied PSG

The self-applied PSG signal used by the sleep technologists to score the self-applied PSG were the EEG signals AF3-E3E4, AF4-E3E4, AF7-E3E4, AF8-E3E4 and the EOG signals E2-AFz, E3-AFz. The EEG signals were filtered using a 0.5-35 Hz bandpass filter and sampled at 200 Hz.

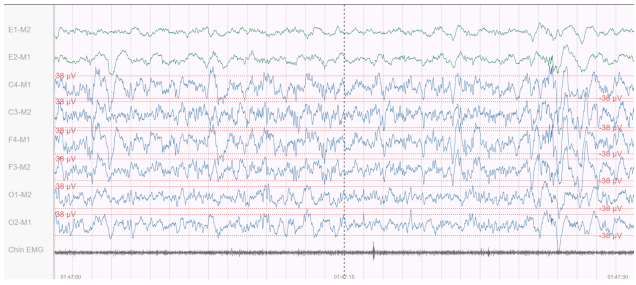

**FIGURE 3** The MicroNyx scoring interface displaying a traditional PSG

The EOG signal was processed through a 0.3-35 Hz bandpass filter and sampled at 200 Hz.

Since the self-applied PSG setup did not include a traditional chin EMG, the E1 signal referenced against the E3 signal, along with the E2 signal referenced against the E4 signal, were used as stand-ins for the EMG signal, with a 10 Hz high-pass filter applied to produce left and right EMG signals, respectively.

Figure 4 shows how the signals from the self-applied PSG were presented to the sleep technologists in the MicroNyx scoring interface.

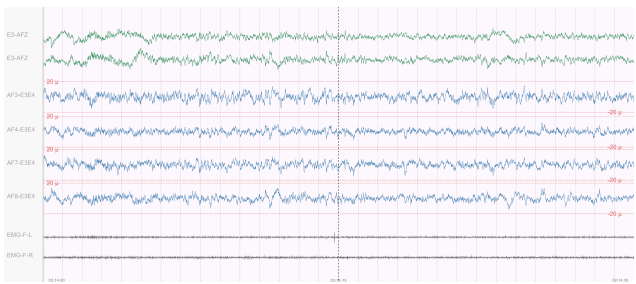

**FIGURE 4** Self-applied polysomnography in the MicroNyx scoring interface

## 2.4 | Reference standard

After the scoring collection, a so-called reference standard was created, a majority-vote scoring for each epoch, which could be used to compare scorings. To have a stable, more reliable hypnogram to compare scorings to than the single-scorer hypnogram, the user scorings for the traditional PSG sessions that did not have recommendations present were used to create a consensus scoring referred to henceforth in this work as the reference standard. This was done to alleviate two major issues. The first issue is that the single-scorer hypnogram was created on different software and is thus not perfectly comparable to the scorings generated on the platform used for this study. The second issue that the reference standard solves is that the scoring of a self-applied PSG is not as validated as a traditional PSG. By creating a consensus scoring for the traditional PSG that can be compared against the scorings for the self-applied PSG, the difference in the scoring of traditional and self-applied PSG is more pronounced. This work uses the reference standard to compare the participant scorings and the recommendations. We refer to sleep technologists' scorings as *accurate* and recommendations as *correct* if they respectively match the reference standard. The words are not considered interchangeable in this work since recommendations cannot be considered 'accurate,' as they are not themselves scorings.

## 2.5 | Recruitment of participants

The study was heavily dependent on the participation of sleep technologists. To recruit sleep technologists as participants in the scoring collection, an email invitation was sent to 49 sleep technologists affiliated with the Sleep Revolution (Arnardottir et al. 2022) project. In addition, the study was advertised at conferences in the European Respiratory Society (ERS) and the European Sleep Research Society (ESRS) in September 2024. An invitation was also emailed to European Society of Sleep Technologists members via a newsletter. In total, 16 sleep technologists across Europe participated in the study.

## 2.6 | Study procedure

Each technologist was instructed to complete a 10-minute (20-epoch) tutorial session in which they were shown how to navigate the scoring interface, how recommendations appeared, and how to score using the MicroNyx interface successfully. After the tutorial, each technologist was directed to complete two scoring sessions, one for PSG and one for self-applied PSG, where they scored exactly one hour of sleep data per session. After those two sessions, the sleep technologists were instructed to wait a week before scoring two additional sessions. The waiting time was instructed in order to prevent familiarity with the recordings.

After successfully completing the four sessions, a link to a short post-study survey was presented to the sleep technologists. The post-study survey aimed to gauge how each sleep technologist perceived the MicroNyx platform for its ease of use and ability to score sleep stages on the scoring interface. The post-study survey questions are provided in appendix A.

The MicroNyx platform ensured that each participant completed one traditional PSG session and one self-applied PSG session in randomized order, followed by a one-week break, after which they repeated the process with a second pair of sessions. Recommendations were presented in only one of the two traditional sessions and one of the two self-applied sessions, ensuring that if recommendations were provided in the first traditional session, they would not be shown in the second traditional session, and the same rule applied to the self-applied sessions. Recommendations are covered in more detail in the following section.

## 2.7 | Analysis

Data analysis was performed using Python and R, with tools chosen to suit the specific requirements of each test. In Python, the Pandas library (McKinney et al. 2010) was used for data preparation and organization. For simpler single-variable analyses, the SciPy library (Virtanen et al. 2020) was employed to conduct hypothesis testing through paired T-tests, assessing significant differences in decision-making time and accuracies under different conditions, assuming normality. When the normality assumption was unmet, the Mann-Whitney test (Mann and Whitney 1947) was applied as a non-parametric alternative. A significance level of $alpha = 0.05$ was used throughout. To investigate the relationship between scoring variables and decision-making time, the R programming language (R Core Team 2023) and the ARTool library (Kay et al. year) were utilized to perform an Aligned-Rank-Transform Analysis of Variance (ART ANOVA). ART ANOVA was chosen for its ability to accommodate the continuous nature of the decision-making time alongside categorical predictors such as recommendation correctness, presentation style, and PSG type, which do not satisfy the assumptions of standard ANOVA. A generalized linear model was applied to analyze scoring accuracy, accounting for the binomial distribution of the dependent variable. This approach enabled the evaluation of categorical predictors' main effects and interactions, such as recommendation correctness, presentation style, and PSG type, while respecting the constraints of binary data.

## 3 | RESULTS

This section presents the results of this study, including the aggregate performance of sleep technologists when scoring traditional and self-applied PSG and the granular effects of recommendations on scoring accuracy and time. The analysis highlights differences between user-level and epoch-level outcomes, focusing on the impact of recommendation correctness and presentation. Cumulatively, the 16 participating sleep technologists completed 64 scoring sessions, producing 9158 individual scorings for the total 240 epochs in the scoring sessions. To create a reference standard, a majority vote approach was taken using the scorings of the 16 participants. This was used to assign a unique label to each epoch. After using a majority-vote system to create the reference standard, the scorings from the original hypnogram used to source the recommendations aligned with the reference standard with 76.23% accuracy; however, due to the intentional 20% error rate, the agreement of the recommended sleep stages with the gold standard dropped to 67.57% for the traditional PSG and 57.64% for the self-applied PSG.

When analyzing the decision-making time, the time per epoch was obtained by calculating the time difference of the creation timestamps of successive scorings. A log transformation was applied to the time-per-epoch data to remove statistical outliers, and an interquartile range filtering with a threshold of 1.5 was used to remove unrealistically long decision-making times. The filtering step identified 598 outliers, with a mean time-per-epoch of 3210.6 seconds and a standard deviation of 77646.7 seconds, indicating that the filtered values deviated significantly from the remaining decision-making times. The remaining data used for the analysis had a mean scoring-per-epoch time of 2.0 seconds and a standard deviation of 1.9 seconds.

## 3.1 | Participants

In total, 16 sleep technologists participated in the study, successfully completing four scoring sessions. Of the 16 sleep technologists, 13 answered the post-participation questionnaire. The questionnaire (see appendix A) showed that most sleep technologists felt confident in their ability to interpret and score signals, with over 84% selecting 6 or 7 on the confidence scale. Two participants (7.7% each) selected 2 or 5, indicating some variation in confidence levels. Of those who answered the questionnaire, all reported that they could easily see the recommendations and that it was easy to see if they were from a human or AI. The sleep technologists were from 11 different countries: Australia, Belgium, Finland, France, Germany, Guatemala, Iceland, Ireland, Italy, Portugal, and Spain.

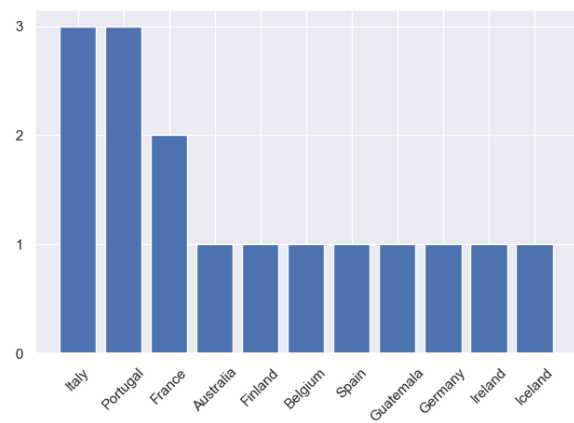

**FIGURE 5** Country breakdown by number of participating sleep technologists.

Of the various recruitment methods, the invitation sent to the relevant members of the Sleep Revolution (Arnardottir et al. 2022) yielded six sleep technologists, and the letter sent to the ESST yielded five. Two sleep technologists participated after hearing about the study from colleagues or during a talk where the study was advertised.

## 3.2 | Aggregate analysis of scoring accuracy and time

To establish a baseline standard for scoring accuracy and time without the influence of recommendations, we filtered the data to exclude sessions that included recommendations from the analysis. When comparing the scoring accuracy of sleep technologists, 11 out of 16 technologists performed better in scoring the traditional PSG sessions than in scoring the self-applied PSG sessions (see in Figure 6 the accuracy achieved by each sleep technologist in both PSGs). Conversely, five sleep technologists scored the self-applied PSG sessions more accurately than the traditional PSG sessions. Overall, sleep technologists demonstrated a tendency for a slightly higher accuracy when scoring the traditional PSG sessions, achieving an accuracy rate of 85.7% compared to 81.0% for the self-applied PSG. The difference in accuracy was, however, not statistically significant (paired t-test: p=0.098, rank-sum: p=0.15).

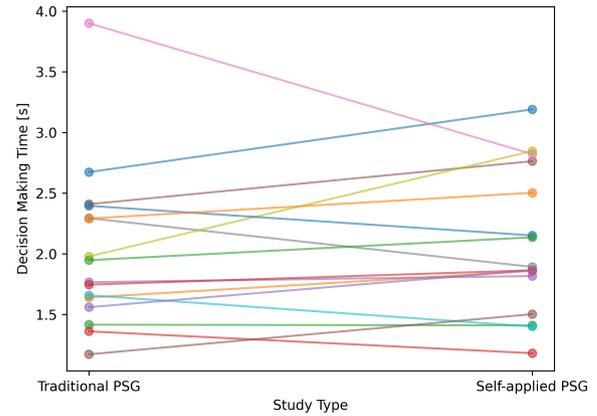

**FIGURE 7** Sleep technologist change in decision-making time between traditional and self-applied PSG. Each line represents one sleep technologist.

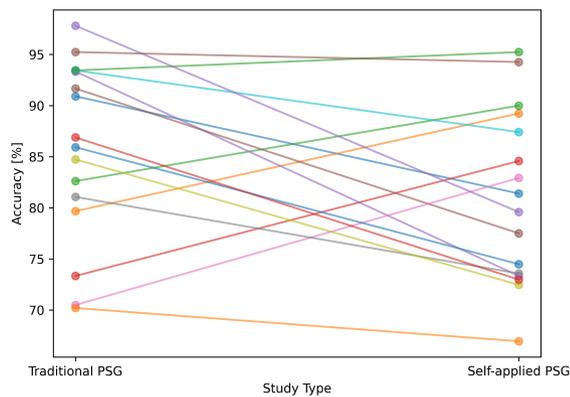

**FIGURE 6** Sleep technologist change in accuracy between traditional and self-applied PSG. Each line represents one sleep technologist.

The decision-making time was analyzed similarly to the scoring accuracy. For the traditional PSG, the average decision-making time was found to be 2.0 seconds, and for the self-applied PSG, 2.0 seconds and is displayed in Figure 7, not statistically significant (paired t-test: p=0.58, rank-sum: p=0.74).

## 3.3 | Epoch-Level effects of recommendations on accuracy

The sessions for both traditional and self-applied PSG were separated based on whether or not they included recommendations, and the scorings were then compared in terms of accuracy with the reference standard. The effect of correct recommendations on the overall accuracy of scorings for all scoring sessions can be seen in the heat map in Figure 8. Notably, the difference between accuracies for the self-applied PSG and the traditional PSG was more dramatic, with the baseline accuracy for self-applied PSG being 81.47%, but decreasing by approximately 3.2% to 78.26% when recommendations were present. The accuracy difference was not statistically significant for the traditional PSG (t-test

p=0.73, rank-sum p=0.73). However, for the self-applied PSG, the difference was found to be statistically significant (T-test p=0.006, rank-sum p=0.006).

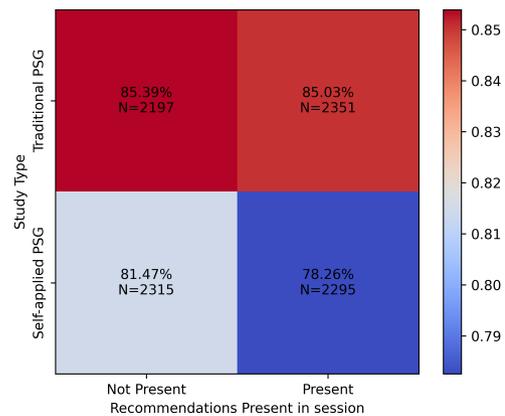

**FIGURE 8** Effect of recommendation presence on scoring session accuracy for traditional vs. self-applied PSG.

When studied further by separating scorings based on their recommendation presence and correctness in Figure 9, the effect of correctness on recommendation becomes clearer. The baseline overall scoring accuracy on an epoch-by-epoch basis without recommendations remained at 85.38% for the traditional PSG sessions and 81.47% for the self-applied PSG sessions.

When faced with incorrect recommendations, sleep technologists assessed the traditional PSG with an accuracy of 82.13%, which is 3.26 percentage points lower than the baseline accuracy of 85.39% achieved when scoring the traditional PSG without any recommendations. This effect is even more pronounced for the self-applied PSG, with the accuracy rate decreasing from the baseline 81.47% achieved when scoring the self-applied PSG without recommendations to 74.20% when faced with incorrect recommendations, making for a 7.27% decrease in accuracy. Correct recommendations had the opposite effect on scoring accuracy, with sleep technologists achieving 90.76% accuracy when scoring traditional PSG epochs featuring an accurate recommendation, resulting in a 5.37% increase in accuracy from the baseline scoring accuracy without recommendations. This positive effect was also observed for the self-applied PSG, where sleep technologists achieved 88% accuracy when presented with correct recommendations, up 6.53% from the 81.47% baseline scoring accuracy for self-applied PSG without recommendations.

**TABLE 1** Generalized linear model linear regression results. The three-factor interaction term was included at first but was not significant. Thus, it was removed from the model.

| | OR | 2.5% | 97.5% | Significance |
|---|---|---|---|---|
| Intercept | 10.492 | 7.543 | 14.595 | * |
| C(Presentation)[T.AI] | 1.215 | 0.776 | 1.901 | |
| C(StudyType)[T.Self-applied] | 0.856 | 0.555 | 1.321 | |
| C(Correctness)[T.False] | 0.379 | 0.244 | 0.588 | * |
| C(Presentation)[T.AI]:C(Study Type)[T.Self-applied] | 0.661 | 0.404 | 1.082 | |
| C(Presentation)[T.AI]:C(Correctness)[T.False] | 1.062 | 0.654 | 1.723 | |
| C(StudyType)[T.Self-applied]:C(Correctness)[T.False] | 0.561 | 0.342 | 0.919 | * |

PSG, incorrect recommendations decreased the scoring accuracy in line with the results from Figure 9. However, the negative impact of incorrect recommendations was not as dramatic for the traditional PSG as it was for the self-applied PSG. The presentation of correct recommendations had a paradoxical effect on accuracy with respect to the study types. For traditional PSG, human recommendations produced a mean accuracy of 90.43%, and AI recommendations produced a mean accuracy of 93.64%. Meanwhile, for self-applied PSG, human recommendations produced an average accuracy of 90.99%, and AI recommendations produced a mean accuracy of 86.77%.

## 3.4 | Epoch-Level effects of recommendations on decision-making time

The effects of recommendations on decision-making time were analyzed using a method similar to the effects on accuracy. When separated based on study type and recommendation presence (see Figure 11), sleep technologists spent 1.9 seconds scoring per epoch on average without recommendations, which rose to 2.0 seconds per epoch when scoring with recommendations. This effect was on the boundary of statistical significance (T-test p=0.041, rank sum p=0.077).

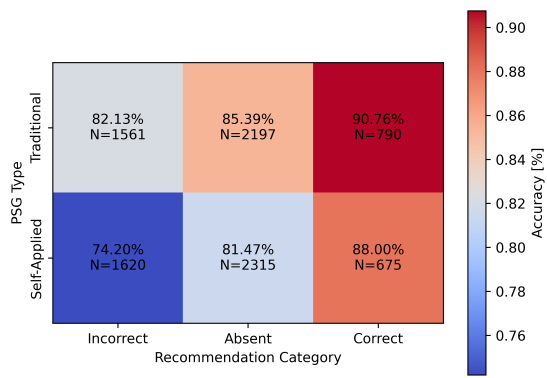

**FIGURE 9** Effect of recommendation correctness on scoring accuracy.

The generalized linear model results can be seen in Table 1, which displays odds ratio (OR) change for each change in variables from the baseline (intercept) where the study type is a traditional PSG, the presentation is human, and the recommendation is correct. The baseline odds ratio is 8.25, meaning that when study type, presentation, and correctness equals the baseline, the accuracy of sleep technologists is 89.18%. The presentation and study type did not affect the scoring accuracy with statistical significance. However, the recommendation accuracy, on its own, had a significant effect on the scoring accuracy, lowering the accuracy of sleep technologists from the 89.18% baseline to 82.54%. The interaction between presentation and study type did not significantly affect the scoring accuracy, with AI recommendations when scoring self-applied PSG lowering the accuracy from the baseline by 10.32% to 78.85%. The final interaction that statistically significantly affected the scoring accuracy was for self-applied PSG when recommendations were incorrect, which lowered the scoring accuracy to 72.34%, or by 16.83%.

When plotted in a three-way line plot (see Figure 10), the interactions from Table 1 become more clear. For both traditional and self-applied

The average time per epoch for self-applied PSG showed a similar trend, with sleep technologists spending 2.0 seconds on average per epoch when scoring without recommendations, which rose to 2.1 seconds when recommendations were introduced. Unlike traditional PSG, this effect was statistically significant (T-test p=0.0036, rank-sum p=0.0039).

Similar to the scoring accuracy, this effect was clearer when scorings were separated based on recommendation correctness and presence (see Figure 12). For both traditional and self-applied PSG, the sleep technologists spent 2.0 seconds on average per epoch when not faced with any recommendations. For incorrect recommendations, the sleep technologists spent 2.2 seconds per epoch on average for both types of PSG. When recommendations are correct, however, the average time-per-epoch decreased by 0.1 seconds for the traditional PSG; however, the decision-making time for self-applied PSG increased by 0.1 seconds on average.

Similarly to the accuracy, ART ANOVA was used to discover which features of the recommendations and their interactions affected the decision-making time. The results of the ANOVA can be seen in Table 2, where the significance levels are marked with one to three asterisks, depending on the significance level. The effect of the presentation alone was insignificant for the decision-making time, and its interaction with the correctness and PSG type did not reach statistical significance. PSG type was found to affect the decision-making time with statistical significance, and its interaction with the correctness of the recommendations had a borderline significant effect on the decision-making time. The correctness was highly statistically significant for the decision-making time. The three-way interaction of the variables had a highly statistically significant effect.

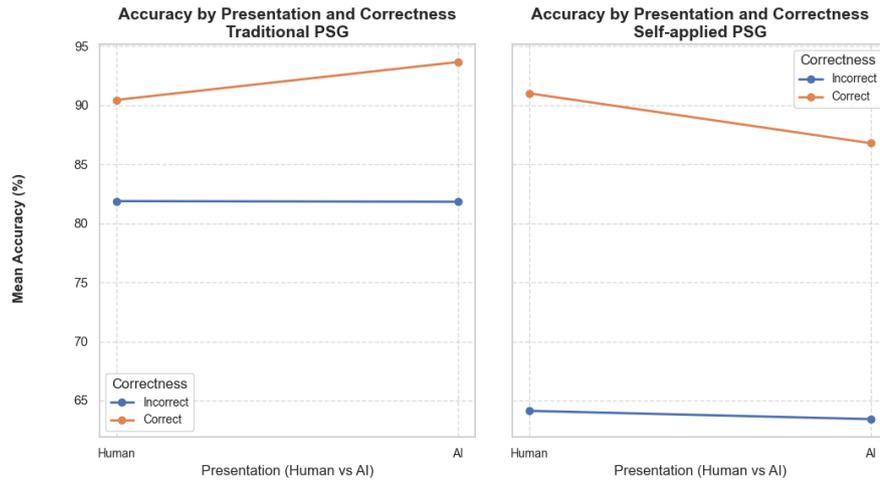

**FIGURE 10** Three-way line plot with grouped comparisons between effects of study type, recommendation presentation, and recommendation correctness on scoring accuracy.

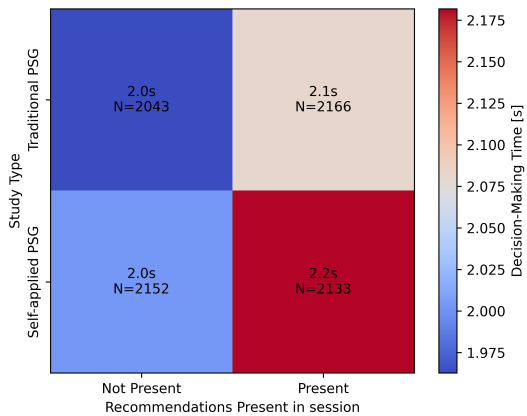

**FIGURE 11** Effect of recommendation presence on decision-making time for traditional vs. self-applied PSG.

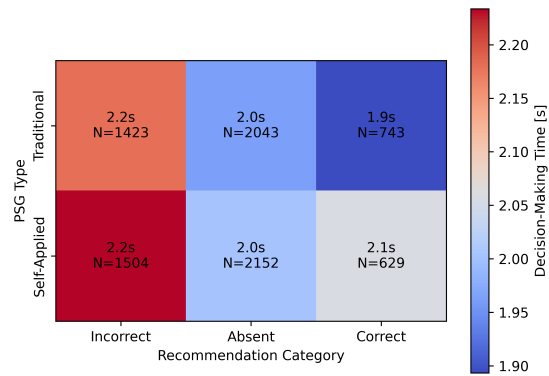

**FIGURE 12** Effect of recommendation correctness on decision-making time.

The three-way interaction between study type, recommendation presentation, and recommendation correctness revealed distinct trends in decision-making time (Figure 13). For traditional PSG, for epochs featuring correct human recommendations, the sleep technologists spent an average of 1.99 seconds per epoch and 1.85 seconds for epochs featuring AI recommendations, or approximately 0.04 seconds shorter when the recommendations were presented as being from AI. For self-applied PSG, correct recommendations displayed a similar trend of sleep technologists taking less time on average to score epochs with AI recommendations (2.0 seconds) vs. human recommendations (2.3 seconds).

**TABLE 2** ART ANOVA results for presentation, study type, and correctness on average decision-making time.

| Effect | Df | Df.res | F value | Pr(>F) |
|---|---|---|---|---|
| Presentation | 1 | 2150 | 0.257 | 0.612 |
| PSG Type | 1 | 2150 | 6.860 | 0.008 ** |
| Correctness | 1 | 2150 | 38.702 | 5.915e-10 *** |
| Presentation:PSG Type | 1 | 2150 | 4.361 | 0.036 * |
| Presentation:Correctness | 1 | 2150 | 0.837 | 0.360 |
| PSG Type:Correctness | 1 | 2150 | 0.391 | 0.531 |
| Presentation:PSG Type:Correctness | 1 | 2150 | 16.193 | 5.917e-05 *** |

Incorrect recommendations showed considerable increases in average decision-making time per epoch, as stated earlier, with sleep technologists spending on average 2.9 seconds scoring epochs with an incorrect human recommendation vs. 2.2 seconds for an incorrect AI recommendation. Self-applied PSG reverses this trend; the sleep technologists spent 2.5 seconds on incorrect human recommendation epochs vs. 3.0 seconds on incorrect AI recommendations.

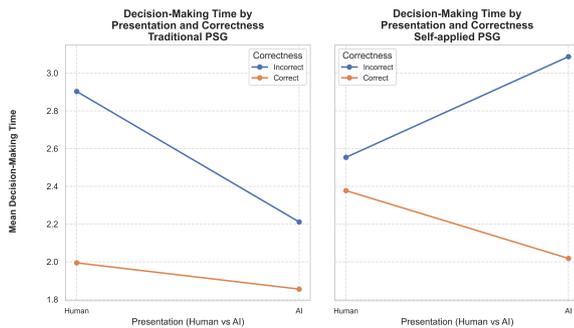

**FIGURE 13** Three-way Line plot with grouped comparisons between the effect of study type, recommendation presentation, and recommendation correctness on decision-making time.

## 4 | DISCUSSION

The main contributions of this work are threefold:

1. We found no significant difference in the scoring accuracy between traditional and self-applied PSG.
2. We found that correct recommendations increased the scoring accuracy for both the traditional and self-applied PSG up to approximately 90% accuracy.
3. We found no evidence for bias toward AI recommendation over Human recommendations when scoring sleep stages.

The findings of this study provide strong evidence for the potential of AI and DSS to enhance sleep stage scoring by improving accuracy and reducing decision-making time. This work contributes significantly to integrating AI-driven and automated scoring systems into the workflows of sleep technologists, paving the way for faster processes and more precise diagnostics.

## Traditional vs. self-applied PSG

When analyzed, no evidence was found that sleep staging accuracy differed for traditional or self-applied PSGs. While the participants were slightly more accurate when scoring traditional PSG epochs, a likely explanation for this is that the majority of sleep technologists participating had not scored a self-applied PSG until in this study. Along with the scoring rules for self-applied PSGs being less defined than for traditional PSGs those two factors are the most likely to affect the scoring accuracy. Our findings align with Rusanen et al. (2023), demonstrating that self-applied PSGs can be reliably scored without additional time or accuracy penalties compared to traditional PSGs. Furthermore, we did not find that the baseline decision-making time differed between the study types, suggesting that the time taken to score self-applied PSGs is not greater than for traditional PSGs.

## Scoring accuracy

The results showed that the baseline scoring accuracy without recommendations was in line with the work of Nikkonen et al. (2024), as sleep technologists achieved a comparable baseline agreement with the prior work exploring the inter-scorer variability in sleep staging. Sleep technologists were more likely to score self-applied PSG epochs incorrectly when presented with incorrect recommendations than with traditional PSG epochs. Both types of PSGs showed reduced scoring accuracy with incorrect recommendations; however, self-applied PSGs suffered a significantly greater decrease relative to its baseline accuracy, with a 7.26 percentage point drop compared to a 3.26 percentage point drop for traditional PSGs. This discrepancy is potentially due to sleep technologists being generally less familiar with self-applied PSG signals, thus deferring the scoring decision to the recommendations. While Rusanen et al. (2023) found that self-applied PSGs tends to suffer from noisier signals than the traditional PSGs or electrode placement issues, this does not apply to this study since the scoring session periods were chosen for their signal quality.

Whether recommendations were presented as being from humans or AI had seemingly no statistically significant effect on the accuracy, indicating a high clinical acquiescence. This finding underscores the potential for seamless integration of AI tools in clinical workflows.

These results have significant clinical implementations, as both the traditional and self-applied PSG scoring show significant improvements in terms of accuracy when correct recommendations are integrated into the scoring process.

## Decision-making time

As Table 2 shows, when epochs with recommendations are analyzed, the study type becomes a significant factor for decision time, with sleep technologists spending an average of 0.2 seconds less time scoring traditional PSG epochs. The correctness was also found to be significant for the decision-making time, with sleep technologists spending 0.3 seconds longer scoring epochs with incorrect recommendations for the traditional PSGs and 0.1 seconds longer for incorrect self-applied PSG recommendations.

The interaction between study type and recommendation presentation also influenced the decision-making time, with sleep technologists spending considerably more time evaluating incorrect AI recommendations for self-applied PSGs than similar recommendations for traditional PSGs. This disparity suggests that, due to their familiarity with traditional PSGs, sleep technologists could more swiftly dismiss incorrect AI recommendations while dedicating additional effort to understanding incorrect human recommendations. Conversely, sleep technologists scored incorrect human recommendations for self-applied PSGs faster than AI recommendations. This indicates that while technologists were more inclined to defer to human recommendations, they subjected AI recommendations to greater scrutiny, reflecting a nuanced approach to clinical acquiescence in the context of novel applications.

The clinical implications for these findings are significant as the reduction in time observed in our results was 0.3 seconds for the traditional PSG epoch, or approximately 13% of the mean decision-making time, meaning that if an eight-hour PSG is scored in two hours, with recommendations, the scoring time will become approximately 103 minutes or a 17 minute gain in time. However, this gain in scoring time is not observed to such a degree for the self-applied PSG epochs, with the time reduction being 0.1 seconds on average, resulting in a 4% speed up, resulting in a scoring time of 114 minutes or a gain of five minutes.

## Study limitations

Despite the promising contributions, the study has several notable limitations. First, the scoring sessions were limited to one hour, or 120 epochs, to reduce participant burden. While practical, this time restriction resulted in a relatively small representation of sleep stages, which may limit the applicability of the findings to full-night studies. Additionally, the one-hour duration was not designed to observe fatigue-related effects that might influence scoring accuracy over extended periods, of interest for future studies, as well as whether scoring is affected by the time of day it is being performed. Second, although the participant pool of 16 sleep technologists provided valuable insights, a larger sample size would be required to ensure statistical power and capture a more comprehensive range of inter-individual variability. While the diversity of participants is a strength, the relatively small number limits the generalizability of some conclusions.

## 5 | CONCLUSION

This study demonstrates that decision support systems have significant potential to affect the scoring accuracy and speed of sleep technologists positively. However, while correct recommendations can make the scoring process more accurate and time-efficient, incorrect recommendations will likely have the opposite effect. Our findings emphasize the critical need for the reliability and correctness of the systems integrated into the workflows of sleep technologists. Additionally, this study provides valuable insights for further research into decision support systems and implementing human-in-the-loop software that incorporates AI into sleep medicine.

Future research should focus on expanding the integration of AI in sleep diagnostics, exploring its application across diverse scoring tasks, and developing systems that empower human experts to deliver more accurate and reliable patient care in less time than currently required. The insights gained from this study pave the way for AI-driven innovations that could revolutionize sleep medicine and enhance patient outcomes. Future research should also examine the long-term effects of AI integration in the scoring process, specifically whether prolonged exposure to AI influences sleep technologists' scoring behaviors or decision-making habits. Future research could involve longer scoring periods or multiple shorter sessions to address the current limitation of insufficient sleep stage variety, incorporating more diverse and representative sleep segments. Furthermore, given time to adjust and learn to recognize the scoring patterns of the recommendations, the positive effect observed for both scoring accuracy and decision-making time could theoretically increase, however that requires study to confirm. Finally, incorporating information on AI uncertainty can increase the DSS's transparency, increasing the trust and clinical acquiescence of the automatic assistance provided to sleep technologists.

This study underscores the need for the accuracy and reliability of DSS tools when incorporated into the scoring process, as incorrect recommendations were shown to impact sleep technologists' accuracy and decision-making time negatively. Furthermore, the results gathered in this study suggest that take-home PSG solutions do not suffer from reduced accuracy in sleep scoring compared to traditional PSG. However, our findings underscore the risks of over-reliance on AI recommendations, particularly in self-applied PSG when incorrect recommendations are involved. If not carefully managed, such reliance can result in systematic errors, potentially compromising diagnostic reliability. To address these challenges, regular calibration and retraining of AI systems and enhanced training for sleep technologists to effectively collaborate with AI are crucial for mitigating risks and ensuring balanced, reliable outcomes.

## AUTHOR CONTRIBUTIONS

CRediT: BH: Conceptualization, Data curation, Formal Analysis, Investigation, Methodology, Project administration, Resources, Software, Supervision, Validation, Visualization, Writing – original draft, Writing – review & editing; AÓ: Investigation, Software; BEÞ: Investigation, Software; HÞH: Investigation, Software; SS: Data curation, Methodology, Validation, Writing – review & editing; HG: Validation; KH: Validation, Writing – review & editing; GMMJ: Writing – original draft, Writing – review & editing; TP: Funding acquisition, Resources, Supervision; ESA: Conceptualization, Funding acquisition, Methodology, Project administration, Resources, Supervision; MÓ: Conceptualization, Formal Analysis, Funding acquisition, Methodology, Project administration, Supervision, Writing – review & editing

## ACKNOWLEDGMENTS

This work would have been impossible to accomplish without the invaluable support of Thomas Penzel and the resident sleep technologists at Charité University Hospital in Berlin. the European Society of Sleep Technologists and its president, Carlos Teixeira, receive special gratitude for allowing us to advertise the study in an email to its members.

## CONFLICT OF INTEREST

Dr Erna Arnardottir reports personal fees from Nox Medical, Jazz Pharmaceuticals, Linde Healthcare, Wink Sleep, Apnimed, Vistor, and ResMed. She is a member of medical advisory boards for Philips Sleep Medicine & Innovation Medical Board and Lilly.

The authors report no other conflicts of interest in this work.

## DISCLOSURE ON THE USE OF GENERATIVE ARTIFICIAL INTELLIGENCE

During the preparation of this article, generative AI tools, specifically ChatGPT (chatgpt.com), were employed to proofread the text and provide suggestions for improvement. The ChatGPT tool was used solely as a refinement aid and was not involved in drafting any part of this work.

Additionally, AI-powered search engines, including Perplexity.ai and Semantic Scholar, were utilized for literature research and sourcing references for this article.

Finally, grammatical corrections were performed using Grammarly (grammarly.com), provided through an educational account by Reykjavik University.

**How to cite this article:** Holm, B., Óskarsson, A., Þorleifsson, B. E., Hafsteinsson, H. Þ., Sigurðardóttir, S., Grétarsdóttir, H., Hoelke, K., Jouan, G. M. M., Penzel, T., Arnardóttir, E. S., and Óskarsdóttir, M.. World of ScoreCraft: Massively Multi-Scorer Online Study on the Effects of Including Decision Support System in Sleep Stages. *Journal of Sleep Research.* 2024;XX(XX):XX–XX.

# APPENDIX

## A  STUDY COMPLETION QUESTIONNAIRE

### Participant Information

1. **What is your name?**
   *(We will not use your personal identity for any analysis or publication, and it will remain completely confidential.)*
   Answer: ____________________________
2. **What is your email?**
   *(If this is not prefilled, please use the same email you used to log into MicroNyx.)*
   Answer: ____________________________
3. **Where are you working?**
   *(Center, City, Country)*
   Answer: ____________________________

### Study Awareness

1. **I heard about this study from:**

   - The invitation email sent to the Sleep Revolution.
   - The talk held by Dr. Erna Sif Arnardóttir at the Sleep Europe conference.
   - The talk held by Dr. Erna Sif Arnardóttir at the ERS conference.
   - Email sent to the members of the ESST.
   - Other: __________________

2. **Are you part of the Sleep Revolution?**

   - Yes
   - No

### User Experience

*This section of the questionnaire is designed to gain insight into the ease of use of the MicroNyx platform.*

1. **The system was easy to use.**
   Strongly Disagree   1   2   3   4   5   6   7   Strongly Agree
2. **I felt that I could reliably read, interpret, and score the signals in the interface.**
   Strongly Disagree   1   2   3   4   5   6   7   Strongly Agree
3. **The scoring recommendations were easy to see.**
   Strongly Disagree   1   2   3   4   5   6   7   Strongly Agree
4. **It was clear which scoring recommendations were from a human, and which were from an AI.**
   Strongly Disagree   1   2   3   4   5   6   7   Strongly Agree
5. **I felt the recommendations were helpful when scoring.**
   Strongly Disagree   1   2   3   4   5   6   7   Strongly Agree

### Additional Feedback

*If you have any further comments, please write them here. We are very happy to hear your feedback.*
Answer: ____________________________